\pdfoutput=1
\documentclass[journal]{IEEEtran}
\usepackage{amsmath,amssymb,amsthm}
\usepackage{graphicx}
\usepackage{booktabs}
\usepackage{array}
\usepackage{multirow}
\usepackage{cite}
\usepackage{xcolor}
\usepackage{tikz}
\usetikzlibrary{arrows.meta,positioning,fit,calc,backgrounds}
\usepackage[hidelinks]{hyperref}

\newtheorem{proposition}{Proposition}
\newtheorem{definition}{Definition}


\newcommand{\NPV}{26}
\newcommand{\PVMW}{8.13}
\newcommand{\Ncases}{1{,}654}
\newcommand{\UARfdiBDD}{95}
\newcommand{\UARfdiEIV}{3.1}

\newcommand{\FRReivLaR}{22}
\newcommand{\LevP}{0.010}
\newcommand{\LevQ}{0.024}
\newcommand{\LevFcst}{0.065}

\newcommand{\PredSD}{0.0008}
\newcommand{\ChiFA}{0.3}
\newcommand{\JdofRatio}{0.99}
\newcommand{\LNRqq}{3.63}

\space
\newcommand{\ADnUnsafe}{143}
\newcommand{\ADnCases}{300}
\newcommand{\ADtelLaR}{87}
\newcommand{\ADtelWLS}{98}
\newcommand{\ADtelProv}{98}
\newcommand{\ADcomLaRnom}{11}
\newcommand{\ADcomLaR}{20}
\newcommand{\ADcomLaRdiff}{9.0~points (6.8--11.3)}
\newcommand{\ADdvMed}{0.11}
\newcommand{\ADdvQ}{0.31}
\newcommand{\ADselChg}{89}
\newcommand{\ADcomWLS}{78}
\newcommand{\ADcomProv}{36}
\newcommand{\ADcomLaRub}{23}
\newcommand{\ADcomWLSub}{80}

\newcommand{\ADdvShare}{71}
\newcommand{\ADnSevere}{39}
\newcommand{\ADsevLaR}{1.7}
\newcommand{\ADsevWLS}{43}
\newcommand{\ETresults}{We ran Haiku and Sonnet on 5~November 2026 with the E2 sampling settings (298 episodes per model and mode; one seed gave no phase-label case) and the two scripts of Table~\ref{tab:checker}.

\emph{Offered the checker, the agents used it, and requiring it changed little.} Haiku called it in 93\% of voluntary episodes (91\% under the note) and Sonnet in all. The mandatory mode raised Haiku's use to 99.7\% ($p<10^{-5}$) but did not reduce violations (Haiku 10 versus 12 of 164 unsafe executions, $p=0.73$; Sonnet 11 versus 14, $p=0.25$; script~A 11). All of the agents' violations came from approvals the checker had verified, except one excursion of its own probe (by 0.01\%); the 23 voluntary approvals without a matching token caused none. Apart from one FDI violation per arm that used the checker (script~B: 52 of 52), all arose under phase-label errors, mostly after the smaller probe (unsafe: 44 of 135 cap vectors verified for the agents at $\epsilon=0.25$, 1 of 119 at $\epsilon=0.5$). The requirement cost Haiku 5 and Sonnet 2 points of safe release, neither significant; for script~B it removed all 72 violations and all release.%
}
\newcommand{\ETdiscussion}{Offered the gate as a tool, both models called it nearly always, even under the note, leaving only the checker's own misses; requiring its token cost 2--5 points of safe release.%
}
\newcommand{\ETabstract}{Given the gate as a tool, they used it in 93--100\% of episodes; requiring it did not reduce violations, which were the checker's own misses.%
}
\space

\begin{document}

\title{Agreement Is Not Assurance: Evidence-Independent Safety Gates for Agentic AI in DER-Rich Distribution Grids}

\author{Qianxue Xia%
\thanks{The author is with the Department of Electrical Engineering and Computer Science, University of Arkansas, Fayetteville, AR 72701 USA (e-mail: qxia@uark.edu).}}

\maketitle

\begin{abstract}
In the emerging safety pattern for large-language-model (LLM) agents in power systems, the agent proposes an action and a physics-based gate decides whether it executes. In distribution grids rich in distributed energy resources (DERs), the gate checks the action against the same state estimate, network records and device settings the agent used, so if that evidence is corrupted, their agreement carries no independent assurance. We formalize this shared-evidence failure: nearly critical DER telemetry lets an adversary shift the estimate while bad-data statistics barely change. We propose an evidence-independent verification (EIV) gate that re-establishes the safety claim at the action's footprint with advanced-metering (AMI) readings anchored to the predicted voltage rise, corroboration points the adversary cannot steer, and a bounded probe. On a hidden-truth IEEE 123-bus testbed with \Ncases{} near-limit requests, stealthy false data made an AC twin gate with bad-data detection approve \UARfdiBDD{}\% of unsafe requests and EIV \UARfdiEIV{}\%. At matched false rejections, the same four readings fed into the state estimator stopped far fewer attacks (32\% versus 8\% of unsafe requests approved). Gate-specific attacks exposed a route through the twin's predicted rise, closed by predicting from commanded set-points; they then got \ADcomLaR{}\% of unsafe requests through EIV and \ADcomWLS{}\% through the AMI-augmented estimator. LLM agents usually verified on their own, but an injected note and phase-label errors led them to unsafe approvals. \ETabstract{}
\end{abstract}


\section{Introduction}
\IEEEPARstart{L}{arge}-language-model agents now plan and execute power-system studies by calling solvers through tool interfaces: optimal power flow and contingency analysis \cite{jin2025gridmind}, violation management with a planning and a validation agent \cite{zhang2025gridagent}, and distribution studies on OpenDSS through dozens of typed tools \cite{liu2026gridorch}. Benchmarks for such agents score evidence-backed findings against hidden evaluators \cite{mylonas2026pab,mylonas2026pabdyn}. As agents move from analysis toward operational recommendations, the community has converged on a division of labor: the agent orchestrates and explains, trusted solvers compute, and a deterministic gate decides what is released. TwinGridShield tests every LLM-proposed action against a network twin before release \cite{rafy2026twingridshield}; a governance layer for a transmission control-room twin enforces tool whitelists, step budgets and operator approval that the model cannot bypass \cite{mylonas2026governance}.

These gates answer an important question: does the action keep the grid safe \emph{if the gate's picture of the grid is right}? They do not answer whether the picture is right. In a DER-rich distribution feeder the picture is assembled from evidence that the agent and the gate share: a state estimate built from sparse SCADA, DER telemetry relayed by vendor clouds and aggregators, and load forecasts; and a network model built from topology, phase and impedance records and DER settings registries. DER telemetry is exposed: a 2025 study disclosed 46 vulnerabilities in inverters and cloud platforms of three major vendors that could allow takeover of inverter fleets \cite{forescout2025sundown}. Distribution records are also often wrong, which is why phase identification from AMI data is an active research topic \cite{olivier2018phase}. If the shared evidence is corrupted, whether maliciously or by accident, the agent proposes an action that the gate approves, because both reason over the same false picture. Agreement between them then reflects a common cause rather than independent confirmation, the failure mode that safety engineering addresses with diversity requirements \cite{iec61508,knight1986independence}.

The gap is acknowledged. TwinGridShield reports that its zero unsafe releases verify conformance to its own rule, that unsafe acceptance rose to 30\% when branch ratings were 20\% optimistic, and that stale topology, AC voltage constraints and adversarial inputs remain future work \cite{rafy2026twingridshield}. PowerAgentBench varies operating points but never corrupts the agent's evidence \cite{mylonas2026pab}. Classical bad-data detection is known to be blind to coordinated false data \cite{liu2009fdi,kosut2011malicious,teixeira2010cyber}, including in distribution systems \cite{deng2019fdidist}. What has not been studied is how physics gates for agentic AI behave when their evidence is corrupted, and what evidence restores their guarantees.

This paper addresses that question for voltage-constrained DER export decisions. Its contributions are:
\begin{enumerate}
\item A formulation of \emph{shared-evidence failure} for action gates, and an analysis showing that the near-criticality of DER telemetry in distribution state estimation lets an adversary shift the estimate, and hence the gate's decision, with little change in chi-square or normalized-residual statistics (Section~\ref{sec:analysis}).
\item An \emph{evidence-independent verification} (EIV) gate that re-establishes the safety claim at the action's footprint with independent-provenance evidence, together with three design rules that the analysis and experiments support: verify the claim rather than the estimate, select corroboration points and predict the action's effect from quantities the adversary cannot bias, and test the model's response with a bounded probe when records may be wrong (Section~\ref{sec:eiv}).
\item A hidden-truth testbed: an IEEE 123-bus feeder with \NPV{} PV systems (\PVMW{}~MW) in OpenDSS, a separate operator twin with a three-phase WLS estimator and bad-data tests, minimum-residual false-data attacks on DER telemetry, six classes of record and setting errors, and a hidden evaluator (Section~\ref{sec:testbed}).
\item A comparison of nominal, margin-based, same-channel, AMI-augmented estimation and independent-evidence gates over \Ncases{} cases, with every gate reading the same physical meters. Under a stealthy attack, an AC twin gate with bad-data detection approved \UARfdiBDD{}\% of unsafe requests and EIV \UARfdiEIV{}\%, at the cost of rejecting \FRReivLaR{}\% of safe requests that all lay within 1\% of the limit; at matched false rejections, the same readings fed into the estimator stopped far fewer attacks. Attacks adapted to each gate, validated on fresh cases, exposed a route through the twin's predicted rise and the amendment that closes it (Section~\ref{sec:results}).
\item LLM-in-the-loop studies with Claude models as proposers. Given provenance-tagged tools and an independent channel, the agents often checked the estimate against AMI on their own, but the check was voluntary: an injected note or a phase-label error led them to unsafe approvals, nearly all of which the enforced gate blocked. Given the gate itself as a tool, they used it in nearly every episode, and making it mandatory did not reduce violations, which were then the checker's own (Section~\ref{sec:llm}).
\end{enumerate}

The gate experiments are proposer-agnostic: every request is evaluated by every gate, so the results hold whether the action comes from an LLM agent, an optimizer, or an aggregator. Section~\ref{sec:llm} then puts LLM agents in the proposer's seat.

\section{Related Work}
\subsection{LLM Agents for Power-System Operations}
Agent frameworks connect LLMs to pandapower, OpenDSS and other solvers so that numerical work is delegated to deterministic tools \cite{jin2025gridmind,zhang2025gridagent,liu2026gridorch}; Grid-Agent adds a validation agent that tests actions in a sandbox \cite{zhang2025gridagent}, and PowerAgentBench evaluates agents with hidden evaluators and evidence logs \cite{mylonas2026pab}. On the safety side, TwinGridShield gates actions with a deterministic DC twin \cite{rafy2026twingridshield}, a governance layer prevents a transmission-twin agent from bypassing approvals \cite{mylonas2026governance}, and NRT-Bench showed that multi-turn red-teaming can drive LLM operator teams in a simulated nuclear control room to lose critical safety functions \cite{lee2026nrtbench}. Threat studies report bad-data injection into LLM inputs in smart grids \cite{li2024risks}, LLMs have classified malicious inverter commands \cite{selim2024llm}, LLM agents have planned coordinated attacks on an industrial testbed \cite{kellerer2026gridstrat}, and the U.S. Department of Energy released a testbed for evaluating agentic AI in operational-technology environments \cite{doe2026stormbreaker}. None of these works corrupts the evidence on which both the agent and its gate rely.

\subsection{Runtime Assurance and Independence}
Gating an untrusted controller with a verified monitor is the idea behind simplex architectures \cite{sha2001simplex} and safety shields for learning agents \cite{alshiekh2018shield}. Both assume that the monitor observes the true state. Safety standards treat common-cause failures with diversity and independence requirements \cite{iec61508}, and experiments with multi-version software showed that independently developed versions fail together more often than independence predicts \cite{knight1986independence}. Indirect prompt injection shows that agents also confuse data with instructions \cite{greshake2023injection,debenedetti2024agentdojo}; our gate sits outside the agent and is unaffected by what the agent believes.

\subsection{False Data, Bad-Data Detection and Active Defenses}
Coordinated false data that lies in the range of the measurement Jacobian passes residual-based bad-data detection \cite{liu2009fdi}; detection and attack trade-offs were analyzed in \cite{kosut2011malicious,teixeira2010cyber}, and distribution-specific attacks in \cite{deng2019fdidist}. Distribution state estimation relies on pseudo-measurements and has low redundancy \cite{baran1994dsse,primadianto2017review,dehghanpour2019survey,vanin2020pmdse}. Active defenses change the physics or the inputs so that attacks become visible: physical watermarking \cite{mo2009watermark} and moving-target defense with D-FACTS perturbations \cite{morrow2012topology,lakshminarayana2021mtd}. Our probe is an action-specific instance of this idea, applied to validate the twin's sensitivity rather than to detect replay. DER-specific attack impacts and mitigations are surveyed in \cite{johnson2019der}.

\section{Problem Formulation}\label{sec:problem}
\subsection{Decision and Safety Model}
Consider a three-phase feeder with node-phase set $\mathcal{N}$ and DERs whose active-power export caps $u$ the operator controls. A request asks to change the caps by $\Delta u$, for example to release curtailment. Let $x$ be the true state, $M$ the true network (topology, phase connections, impedances, regulator taps, capacitor states and DER control settings), and $v^{+}(\Delta u)=v(\Phi(x,\Delta u;M))$ the true post-action voltage magnitudes. The request is safe if $V_{\min}\le v^{+}_j\le V_{\max}$ for all $j\in\mathcal{N}$; we use $[0.95,1.05]$~pu following ANSI C84.1 Range~A \cite{ansic84}.

The operator holds a twin $\hat M$ built from records $\mathcal{R}$ and an estimate $\hat x=\mathrm{SE}(z_{\mathcal P};\hat M)$ from measurements $z_{\mathcal P}$ on primary channels $\mathcal P$ (SCADA, DER telemetry, load forecasts). The twin predicts
\begin{equation}
\hat v^{+}_j(\Delta u)=\hat v_j+\Delta\hat v_j(\Delta u),\quad \hat v_j=|\hat x_j|,\label{eq:pred}
\end{equation}
where $\Delta\hat v_j$ is the change computed by an AC power flow on $\hat M$ initialized at the estimated injections. A \emph{gate} $G(\Delta u,\mathcal E)\in\{0,1\}$ maps the request and an evidence set $\mathcal E$ to approve or reject. The nominal gate is
\begin{equation}
G_{\mathrm{nom}}=\mathbf{1}\big[\mathrm{BDD}(z_{\mathcal P};\hat M)\ \wedge\ \max_j \hat v^{+}_j\le V_{\max}-\delta\big],\label{eq:gnom}
\end{equation}
with bad-data detection (BDD) by the chi-square test and the largest normalized residual (LNR) \cite{abur2004se}, and margin $\delta\ge0$.

\subsection{Provenance and Shared-Evidence Failure}
Each evidence item $e$ has a provenance $\pi(e)\subseteq\mathcal C\cup\mathcal R$: the channels and records it depends on. The nominal gate's certificate has provenance $\mathcal P\cup\mathcal R$, the same as the agent's own view.
\begin{definition}[Shared-evidence failure]
A corruption $\kappa$ of channels $C\subseteq\mathcal P$ or records $R\subseteq\mathcal R$ causes a shared-evidence failure for request $\Delta u$ if $G(\Delta u,\mathcal E_\kappa)=1$ while $v^{+}(\Delta u)$ violates the limits.
\end{definition}
We consider two corruption families. \emph{Channel corruption}: an adversary controls the telemetry of a channel (here the DER cloud or an aggregator), knows the operator's model and measurements, and cannot alter channels it does not control (SCADA, AMI). \emph{Record corruption}: non-adversarial errors in phase labels, impedances, DER settings, reported regulator taps, capacitor status or switch status. The intent behind a corruption is not identifiable from the evidence; a malicious parameter change and an accidental one look the same, so the gate must handle both.

\subsection{Metrics}
For each gate we report the unsafe-approval rate (UAR, approvals among truly unsafe requests), the severe UAR (requests whose true maximum exceeds 1.055~pu), the false-rejection rate (FRR, rejections among truly safe requests without corruption), and the verification cost (AMI reads and probes).

\section{Why Shared-Evidence Gates Fail}\label{sec:analysis}
Linearize the WLS estimator at the operating point. With Jacobian $H$, weights $W$, gain $K=(H^\top WH)^{-1}H^\top W$ and residual operator $S=I-HK$, an additive corruption $a$ supported on rows $C$ shifts the estimate by $K_Ca_C$ and the residual by $S_Ca_C$.

\begin{proposition}[Decision invariance]\label{prop:inv}
If $S_Ca_C=0$, every residual-based statistic is unchanged and, to first order, $G_{\mathrm{nom}}$ returns on the corrupted evidence the decision it would return on uncorrupted evidence from the state $x+K_Ca_C$. More generally the chi-square statistic acquires noncentrality $\lambda=\|W^{1/2}S_Ca_C\|^2$ while the estimate of node-phase $j$ is biased by $b_j=g_j^\top K_Ca_C$, with $g_j=\nabla_x|x_j|$. The least noncentrality that produces bias $b$ is $\lambda^\star=b^2/(q_j^\top \Sigma^{+}q_j)$ with $q_j=K_C^\top g_j$ and $\Sigma=S_C^\top WS_C$, and $\lambda^\star=0$ whenever $q_j$ has a component in the null space of $\Sigma$.
\end{proposition}
The proof follows from linearity: residual statistics are functions of $Sz$, and $S(z+a)=Sz+S_Ca_C$; the bias-noncentrality bound is a Cauchy--Schwarz argument on the generalized Rayleigh quotient. The practical content is in the leverage $\ell_i=[HK]_{ii}$: a row with $\ell_i\approx1$ is nearly critical, and changes to it barely reach the residual. In our estimator (Section~\ref{sec:testbed}) the median $1-\ell_i$ of DER active and reactive power rows is \LevP{} and \LevQ{}, against \LevFcst{} for load pseudo-measurements. DER telemetry therefore steers the estimate almost for free, and the attack in Section~\ref{sec:testbed} exploits exactly this.

Record corruption acts differently: it changes $\hat M$, and with it both $\hat v$ and the predicted change $\Delta\hat v$. The estimator absorbs part of the inconsistency into soft pseudo-measurements, so bad-data tests catch gross record errors (switch status, a mis-stated 600~kvar bank) but not small ones (one phase label, a lateral impedance, one inverter's volt-var setting).

\section{Evidence-Independent Verification}\label{sec:eiv}
\subsection{Design Rules}
EIV adds evidence with provenance disjoint from the corruptible inputs and uses it to re-establish the safety claim where the action acts. Three rules follow from the analysis and are tested in Section~\ref{sec:results}.

\emph{R1: verify the claim, not the estimate.} A consistency test between independent readings and $\hat v$ detects a bias only when it is large relative to the readings' accuracy, and an adversary can confine the bias to the few node-phases that matter. EIV instead recomputes the safety claim from the independent reading,
\begin{equation}
y_j+\Delta\hat v_j(\Delta u)\le V_{\max}+\tau\quad \forall j\in\mathcal S_B,\label{eq:anchor}
\end{equation}
where $y_j$ is an AMI voltage reading at corroboration point $j$, $\mathcal S_B$ is the set of $B$ points, and $\tau$ is a tolerance.

\emph{R2: select and predict from quantities the adversary cannot bias.} The points are chosen from the action's footprint, ranked by the predicted rise $\Delta\hat v_j(\Delta u)$, which depends on the action and the records but only weakly on the corrupted estimate. Ranking by the predicted level $\hat v^{+}_j$ would let the adversary push the critical point out of the set. The rise itself must be computed from the commanded change, starting the requesting units from the caps the operator set rather than from their telemetry-estimated output: a twin that adds the request to the reported output and caps the result at nameplate lets false telemetry hide the unit's headroom and shrink $\Delta\hat v$ (Section~\ref{sec:adaptive}).

\emph{R3: test the model's response when records may be wrong.} Channel corruption leaves $\Delta\hat v$ nearly intact, so R1 suffices; record corruption corrupts $\Delta\hat v$ itself. A bounded probe applies a fraction $\epsilon$ of the request and compares the observed change at the corroboration points, whose static meter bias cancels, with the predicted change.

\subsection{Guarantees}
\begin{proposition}[Bounded undetected violation]\label{prop:bound}
Let $y_j=v_j+e_j$ with $\pi(y_j)\cap(C\cup R)=\emptyset$, and let the twin's predicted change satisfy $|\Delta v_j-\Delta\hat v_j|\le\eta_j$. If \eqref{eq:anchor} holds at $j$, then $v^{+}_j\le V_{\max}+\tau-e_j+\eta_j$, independently of the corruption of $\mathcal P$.
\end{proposition}
Under channel corruption, with the rise predicted from commanded set-points, $\eta_j$ is the twin's ordinary prediction error (standard deviation \PredSD{}~pu here) plus whatever the adversary can induce through nonlinear twin effects, so undetected violations at corroborated points are bounded by the meter error and that induced error. The bound applies only to corroborated points, which is why R2 matters:
\begin{proposition}[Selection]\label{prop:sel}
If $\mathcal S_B$ is a function of $\hat v$, an adversary that can bias $\hat v_{j^\star}$ by $b$ removes $j^\star$ from $\mathcal S_B$ whenever $b$ exceeds the gap between $\hat v^{+}_{j^\star}$ and the $B$-th largest $\hat v^{+}$. If $\mathcal S_B$ depends only on $\Delta\hat v(\Delta u;\hat M)$, computed from commanded set-points, channel corruption cannot change it to first order.
\end{proposition}
The first-order statement is local: it holds where $\Delta\hat v$ is smooth in the estimated state, and an adversary can still move $\Delta\hat v$ through nonlinear twin effects such as inverter volt-var breakpoints; Section~\ref{sec:adaptive} measures how far.
For the probe, let $q_j$ be the difference between observed and predicted changes. Under a correct model $q_j\sim\mathcal N(0,2\sigma_r^2+\eta^2)$, where $\sigma_r$ is the per-read noise. A sensitivity error of relative size $\kappa$ shifts the matched-filter statistic $T=\Delta\hat v^\top q/(s\|\Delta\hat v\|)$ by $\kappa\|\Delta\hat v(\epsilon\Delta u)\|/s$, so power grows with $\epsilon$ and with the footprint's sensitivity. The probe size is chosen so that the probe is safe even if the true sensitivity is $\rho$ times the predicted one: $\epsilon=0.5$ if $y_j+\rho\,\Delta\hat v_j(0.5\Delta u)\le V_{\max}-2\sigma$ for all $j\in\mathcal S_B$, else $\epsilon=0.25$, with $\rho=1.5$.

\subsection{Gate and Agent Integration}
\begin{figure}[t]
\centering
\includegraphics{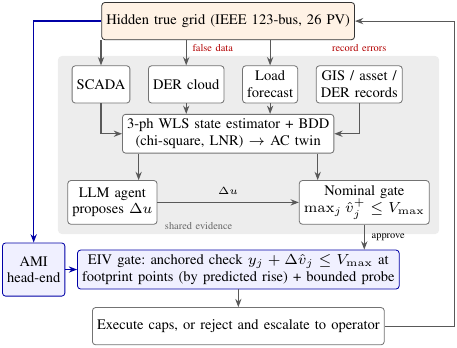}
\caption{Shared versus independent evidence. The agent and the nominal gate both depend on the primary channels and the records (gray); the EIV gate re-establishes the safety claim with AMI readings and a bounded probe whose provenance is disjoint from them.}
\label{fig:arch}
\end{figure}

The EIV gate approves a request only if the nominal gate approves, \eqref{eq:anchor} holds at all $B$ points, and, when the probe is enabled, neither the matched-filter nor the maximum statistic of $q$ exceeds its threshold (family-wise level 1\%). Otherwise it rejects and escalates to a human operator with the failed check. The gate sits outside the agent: the agent may request corroboration through its tools, but it cannot approve on its own evidence, and every tool output carries provenance tags so that the agent's report can be audited against the gate's evidence. Fig.~\ref{fig:arch} summarizes the architecture.

\section{Testbed and Experimental Design}\label{sec:testbed}
\subsection{Hidden True Grid}
\begin{figure}[t]
\centering
\includegraphics[width=0.88\columnwidth]{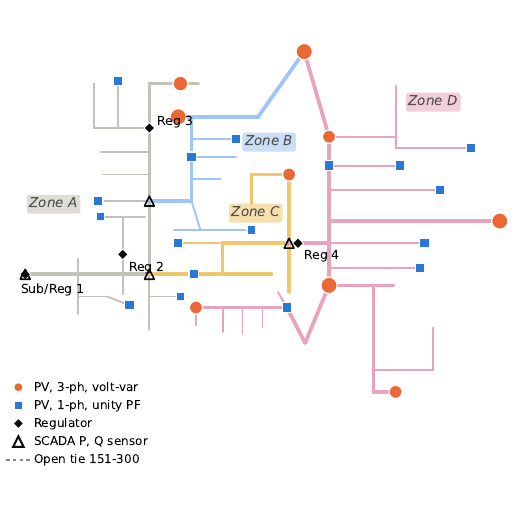}
\caption{IEEE 123-bus test feeder with the PV fleet, regulators and SCADA flow sensors.}
\label{fig:feeder}
\end{figure}

The true grid is the IEEE 123-bus feeder \cite{kersting2001feeders} in OpenDSS \cite{dugan2011opendss} with the following changes. Loads are constant-power, wye-connected and single-phase per node (delta and three-phase loads are split); the unloaded 480~V transformer is removed; closed switches have 1~m$\Omega$ impedance and open ties are disabled lines; regulator leakage reactance is 0.5\%; the substation regulator holds 1.02~pu without line-drop compensation; and the 600~kvar bank at bus 83 is off for midday operation. We add \NPV{} PV systems with \PVMW{}~MW of nameplate capacity (Fig.~\ref{fig:feeder}): nine three-phase units with IEEE~1547-2018 Category~B volt-var control \cite{ieee1547} and seventeen single-phase legacy units at unity power factor, each behind a short service line.

Each case samples a midday operating point: a load level uniform in $[0.30,0.65]$ of peak with lognormal diversity, true loads that differ from the forecasts by 30\% (standard deviation), irradiance uniform in $[0.65,1]$, and a fleet export cap set by bisection so that the pre-action maximum voltage matches a draw from $U(1.025,1.044)$~pu. A group of two to five PVs in one zone, curtailed to at most 20\% of available power, requests release. The released amount is set by bisection on the true grid so that the true post-action maximum is drawn from $U(1.0505,1.065)$ (unsafe) or $U(1.04,1.0495)$ (safe) with equal probability. Requests therefore sit near the limit by design, which makes the test demanding. Regulator taps are settled before the action and held during it.

\subsection{Operator View}
The operator twin is a separate OpenDSS model built from records. Table~\ref{tab:meas} lists the measurement channels. The estimator is a three-phase WLS in rectangular coordinates, solved by Gauss--Newton with QR factorization of the weighted Jacobian; zero-injection nodes are virtual measurements and the substation provides balanced angle references. On clean data its chi-square false-alarm rate was \ChiFA{}\% at the 1\% level, the mean ratio of the weighted residual sum to its degrees of freedom was \JdofRatio{}, the 99th percentile of the LNR was \LNRqq{} (threshold 4.0), and the twin's predicted feeder maximum differed from the truth with standard deviation \PredSD{}~pu.

\begin{table}[t]
\caption{Measurement channels and their provenance}\label{tab:meas}
\centering\footnotesize
\begin{tabular}{@{}l >{\raggedright\arraybackslash}p{3.2cm} >{\raggedright\arraybackslash}p{1.95cm}@{}}
\toprule
Channel & Quantities & Noise (1$\sigma$)\\
\midrule
SCADA & $|V|$ at substation and regulator outputs; per-phase P, Q at four automated switches & 0.3\%; 1\% + 0.5~kW\\
DER cloud & P, Q, $|V|$ of each PV & 1\% of rating; 0.3\%\\
Forecast & load P, Q (pseudo-measurements) & 30\%\\
Records & topology, phases, impedances, taps, capacitor and DER settings & --\\
\midrule
AMI & $|V|$ at load and PV service points, on-demand reads (independent channel) & 0.3\% static + 0.1\% per read\\
\bottomrule
\end{tabular}
\end{table}

\subsection{Corruptions}
\emph{FDI (DER cloud).} The adversary controls all DER telemetry rows, knows the model and the measurements, and solves
\begin{align}
\min_{a_C}\ & \|W^{1/2}S_Ca_C\|^2+\gamma\|a_C\|^2_{W}\nonumber\\
\text{s.t.}\ & \hat v^{+}+J_VK_Ca_C\le V_{\max}-\mu,\quad |a_C|\le\bar a,\label{eq:attack}
\end{align}
re-linearizing at the attacked estimate up to four times. The target margin $\mu\in\{0.2,0.7,1.2\}\%$ is chosen per case and per gate to maximize approval, which gives each gate its worst case among these strategies. \emph{FDI (aggregator)} restricts the adversary to the requesting group's telemetry. The record and setting errors are: a single-phase PV recorded on the wrong phase; the impedance of the lateral feeding a requesting PV 60\% higher than recorded; a requesting unit's volt-var function disabled while the registry shows it enabled; a regulator tap reported one step low; a 50~kvar bank recorded off while on; and a section transferred to the tie at bus 151 while the records show the normal configuration.

\subsection{Gates}
We compare: \emph{Twin}, the AC twin check \eqref{eq:gnom} with $\delta=0$ and no BDD (an AC analogue of TwinGridShield); \emph{Twin+BDD}; \emph{Margin} with $\delta\in\{0.5,1\}\%$; \emph{Same-ch.}, which re-polls four DER-cloud voltages in the footprint and applies the consistency test, and \emph{Same-ch.\ anch.}, which applies \eqref{eq:anchor} to them; \emph{EIV-L}, a consistency-only test of four AMI readings against $\hat v$ (sum and maximum of standardized residuals, family-wise level 1\%); \emph{EIV-La}, which adds the anchored check \eqref{eq:anchor} with $B=4$ and $\tau=0$; \emph{EIV-R}, the probe test alone; and \emph{EIV-LaR}, the anchored check followed by the probe. To ask whether the same readings would serve as well inside the estimator, we add \emph{AMI-WLS}, the WLS estimate with the four AMI rows added (variance $0.003^2+0.001^2$), recalibrated chi-square and LNR tests and the AC check on the new estimate, and a \emph{down-weighted AMI-WLS} whose DER-cloud rows have their variance multiplied by $\kappa$ in every case. AMI-WLS uses a rank-four update of the converged estimate; full nonlinear re-solves of 474 draws changed its predicted maximum by at most 0.02\% and its decision in 0.2\% of them. Every gate that reads meters does so only after the nominal gate approves, uses the same $B$ node-phases ranked by predicted rise (R2), and sees the same physical readings: in each of 20 repetitions per case, a meter's static bias and per-read noise come from one generator keyed by case, repetition, meter and measurement stage, so the bias persists across a meter's pre-probe and post-probe readings and every gate that reads that meter at that stage receives the same value. When gates are compared at matched false rejections, $\tau$, the recalibrated thresholds and $\kappa$ are set on development cases (even case index) and results are reported on held-out cases (odd index), with case-level bootstrap intervals.

\section{Results}\label{sec:results}

\subsection{Calibration and Clean Operation}
Without corruption the nominal gate is accurate. Twin+BDD approved 3\% of unsafe requests, none with a violation above 0.5\%, and rejected 3\% of safe ones (Table~\ref{tab:main}). Margins are costly because every safe request lay within 1\% of the limit: a 0.5\% margin rejected 51\% of safe requests and a 1\% margin 100\%. EIV-La rejected 21\%, concentrated next to the limit: 56\% of safe requests within 0.25\% of it, 24\% between 0.25\% and 0.5\%, and 6\% and 1\% in the next two 0.25\% bands.

\subsection{Stealthy False Data Defeats Shared-Evidence Gates}
Without attack, Twin+BDD would have rejected 99\% of the truly unsafe requests in the FDI set. With the minimum-residual attack aimed 0.2\% below the limit, the nominal gate approved 94\% of them while both bad-data tests passed. The median increase of the chi-square statistic was 0.9 (90th percentile 11.1), small against the distance of about 30 between its typical clean value and the threshold, and the median largest change in a reported DER voltage was 0.53\%. Fig.~S1 in the supplement shows one case. Against the attacker's best target margin per case, Twin+BDD approved 95\% of unsafe requests (95\% Wilson interval 90--97\%) and 83\% of those with violations above 0.5\%. Margins did not help against an attacker who knows them: a 0.5\% margin approved 71\%, and a 1\% margin approved 22\% while rejecting 100\% of safe requests. An adversary limited to the requesting aggregator's telemetry was easier to see: Twin+BDD approved 39\%, although the twin without bad-data tests still approved 97\%.

\begin{table*}[t]
\caption{Unsafe requests approved (\%) under each corruption, severe approvals under FDI, and cost on clean cases (all 1{,}654 cases). Worst case over three attacker target margins; $B=4$ corroboration points selected by predicted rise; 20 paired reading draws per case.}\label{tab:main}
\centering\footnotesize
\begin{tabular}{@{}lrrrrrrrr|rrr@{}}
\toprule
 & \multicolumn{8}{c|}{Unsafe approved (\%)} & FDI & \multicolumn{2}{c}{Clean}\\
Gate & FDI cloud & FDI aggr. & Phase & Line & Volt-var & Tap & Capacitor & Switch & $>$0.5\% & FRR (\%) & Reads\\
\midrule
Twin (AC, no BDD) & 100 & 97 & 43 & 24 & 12 & 19 & 0 & 44 & 100 & 2 & 0.0 \\
Twin + BDD & 95 & 39 & 42 & 24 & 12 & 14 & 0 & 0 & 83 & 3 & 0.0 \\
Margin 0.5\% & 71 & 0 & 13 & 0 & 0 & 0 & 0 & 0 & 37 & 51 & 0.0 \\
Margin 1\% & 22 & 0 & 0 & 0 & 0 & 0 & 0 & 0 & 3 & 100 & 0.0 \\ \midrule
Same-channel & 94 & 35 & 41 & 24 & 12 & 12 & 0 & 0 & 82 & 3 & 0.0 \\
Same-channel, anchored & 82 & 21 & 29 & 14 & 3 & 1 & 0 & 0 & 60 & 18 & 0.0 \\ \midrule
EIV-L (consistency) & 58 & 32 & 40 & 24 & 12 & 11 & 0 & 0 & 8 & 4 & 2.0 \\
EIV-La (anchored) & 3 & 3 & 28 & 12 & 3 & 1 & 0 & 0 & 0 & 21 & 2.0 \\
EIV-R (probe) & 94 & 39 & 21 & 23 & 12 & 14 & 0 & 0 & 83 & 3 & 3.9 \\
EIV-LaR & 3 & 3 & 12 & 10 & 3 & 1 & 0 & 0 & 0 & 22 & 3.5 \\
\midrule
$n_{\text{unsafe}}$ & 147 & 98 & 98 & 95 & 73 & 78 & 47 & 32 & 35 & 148 &  \\
\bottomrule
\space
\end{tabular}
\end{table*}

\subsection{Independence, Anchoring and Selection}
Re-polling the compromised channel did not help: the same-channel consistency test approved 94\% of unsafe requests and its anchored variant 82\%, since the adversary answers the re-poll with the same falsified values. Independent readings helped only when used to recompute the claim (R1). The consistency-only test EIV-L approved 58\%, because the adversary confines its bias to a few node-phases where a 0.3\% meter error hides it, whereas the anchored EIV-La approved 3.2\% and none of the severe cases. As Proposition~\ref{prop:bound} predicts, the probe adds nothing against channel corruption (EIV-R: 94\%), and EIV-LaR approved 3.1\% (95\% Wilson interval 1.3--7.3\%) with no severe approvals. Selection mattered as Proposition~\ref{prop:sel} predicts: ranking points by the estimated post-action level, which the adversary can lower, raised EIV-La's unsafe approvals under FDI from 3.2\% to 15\%, and widening the footprint to every phase at the affected buses did not reduce phase-label approvals at $B=4$ (supplement, Table~S6).

\subsection{Anchoring Versus Estimation}
\begin{figure*}[t]
\centering
\includegraphics[width=\textwidth]{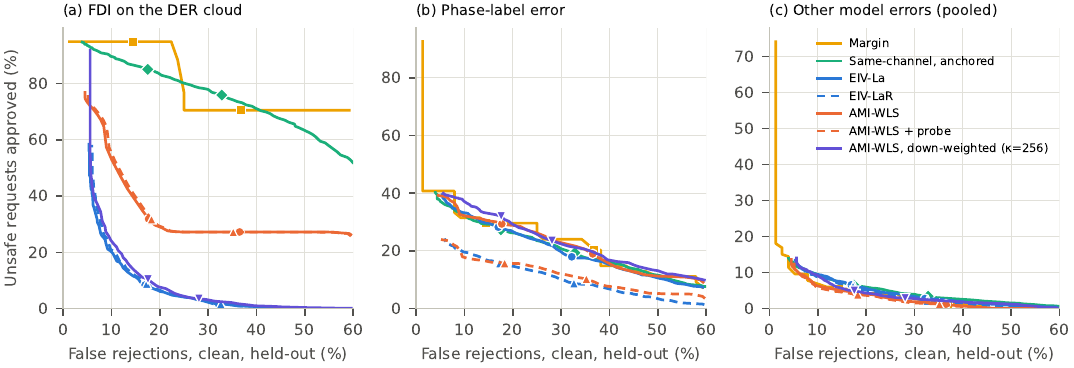}
\caption{Safety-availability frontiers on held-out cases for gates that read the same four AMI meters (curves: tolerance $\tau$ varied; markers: operating points calibrated to 10\% and 20\% clean false rejections on development cases). AMI-WLS absorbs the readings into the estimate; the down-weighted estimator multiplies the variance of every DER-cloud row by $\kappa=256$. (a) FDI on DER-cloud telemetry, worst case over three attacker target margins. (b) Phase-label errors. (c) Line, volt-var, tap, capacitor and switch errors pooled.}
\label{fig:frontier}
\end{figure*}

Fig.~\ref{fig:frontier} compares, on held-out cases, the gates that read the same four meters. At the operating point calibrated to 10\% clean false rejections on development cases, EIV-La approved 8.4\% of unsafe requests under FDI (95\% interval 6.0--10.9\%) and AMI-WLS 31.9\% (24.8--39.0\%); the paired difference was 23.5 points (17.6--29.5), and 25.6 points (18.3--32.7) at the 20\% point. The estimator absorbs the readings into the attacked state instead of letting them overrule it: the four AMI rows compete with nearly critical DER rows of similar accuracy, and the augmented estimate removed a median 16\% of the attack's level shift at the metered nodes (interquartile range 12--21\%). Its recalibrated bad-data tests flagged most attacked draws, but attacks that hid the violation deeply passed them, so AMI-WLS levels off at about 27\% of unsafe approvals however strict its tolerance. Ranking the meters by estimated level hurt it as it hurt the anchored check (31.9\% to 52\%), and with 16 meters it still approved 15\%, about what EIV-La approved with one.

Down-weighting every DER-cloud row ($\kappa=256$, chosen on development cases) closed most of the gap: 10.4\% at the 10\% point, a paired difference from EIV-La of 2.1 points ($-1.0$ to 5.3), and 3.5\% against 1.7\% at the 20\% point. That estimator needs a provenance weighting policy, a tuned $\kappa$, recalibrated thresholds and a re-estimate per request (0.18~s here, against 0.2~ms for the anchored check), and its decision rests on an estimate an auditor must trust, whereas the anchored check is one inequality per meter. Estimation did better where the corrupted rows were fewer: when only the requesting aggregator's telemetry was attacked, the down-weighted estimator approved 8.1 points fewer unsafe requests than EIV-La (4.9--11.5), and under volt-var setting errors AMI-WLS approved 5.8 points fewer (1.7--10.8). Anchoring's advantage grew with meter error relative to DER telemetry: with 0.1\% static AMI error both approved under 1\% at the 10\% point, with 0.2\% 2.8\% against 10.9\%, and with 0.5\% 25\% against 66\%. The development-calibrated points gave 15--18\% false rejections on held-out cases for every gate, with no significant difference between gates.

\subsection{Adaptive Attacks}\label{sec:adaptive}
The attacks above were designed against the nominal gate. We then let the adversary target each gate. It controls the same rows within the same bounds and knows every gate and its frozen settings, hence which meters a gate will read, but not the readings: it scores a common pool of candidate attacks per case with surrogate readings built from the clean estimate and its own noise samples, picks the best candidate for each gate, and the choice is fixed before the evaluation readings are drawn. The pool (18 candidates, the same for every target) contains minimum-residual attacks at three target margins, attacks that also penalize the predicted rise at the meters the gate would read, and attacks on the requesting units' reported output. The attack recipe was developed on existing cases and frozen before a fresh validation block was opened (\ADnUnsafe{} unsafe of \ADnCases{} new FDI cases).

Development exposed the last family. The twin added the request to the units' telemetry-estimated output and capped the sum at nameplate, so reporting an output close to nameplate hid the units' headroom: the twin predicted a fraction of the true rise, the response stayed linear over the probe's range, and the attack stayed far below the bad-data thresholds. On the validation block, the attack chosen for each gate made EIV-LaR approve \ADtelLaR{}\% of unsafe requests, AMI-WLS \ADtelWLS{}\% and the down-weighted estimator \ADtelProv{}\%. Every gate that trusts the twin's predicted rise inherits this route, which R2 now closes by predicting from commanded caps; on clean cases the change moved the predicted rise by at most 0.004\%, so the frozen settings were kept.

With the commanded basis, the attack chosen for EIV-LaR raised its unsafe approvals from \ADcomLaRnom{}\% (attack chosen for the nominal gate) to \ADcomLaR{}\% (difference \ADcomLaRdiff{}), by moving the predicted rise at the selected meters through nonlinear twin effects: by a median \ADdvMed{}\% and a 90th percentile of \ADdvQ{}\%, with at most one of the four meters changed in \ADselChg{}\% of cases. We had predicted before the development run that at least 90\% of these attacks would move the rise by at most 0.2\%; \ADdvShare{}\% did. The attack chosen for AMI-WLS made it approve \ADcomWLS{}\% and the one chosen for the down-weighted estimator \ADcomProv{}\%; even an attacker that could see the readings and pick the best candidate in the pool reached \ADcomLaRub{}\% against EIV-LaR, against \ADcomWLSub{}\% against AMI-WLS (supplement, Table~S9). Of the \ADnSevere{} requests with violations above 0.5\%, the targeted attacks got \ADsevLaR{}\% through EIV-LaR and \ADsevWLS{}\% through AMI-WLS.

\subsection{Record and Setting Errors}
Record errors caused fewer unsafe approvals than the adversary but were harder to remove. Phase-label errors were the most dangerous: Twin+BDD approved 42\% of unsafe requests and 44\% of the severe ones, because the twin places the unit's injection on the wrong phase and AMI meters inherit the same phase labels. The anchored check reduced this to 28\%, the probe alone to 21\%, and both together to 12\%; EIV-LaR also rejected 30\% of the safe requests in these cases, which are correct escalations of a wrong model. Of the unsafe draws EIV-LaR still approved, 71\% had a selected meter whose true rise exceeded the prediction by more than 0.2\%; most were probed at $\epsilon=0.25$, where the discrepancy reached a median of only 1.9 noise standard deviations, and 27\% had no selected meter above the limit. For lateral impedance errors unsafe approvals fell from 24\% to 10\%, for disabled volt-var from 12\% to 3\%, and for tap misreports from 14\% to 0.6\% (EIV-La). Bad-data detection caught the gross errors: no gate with BDD approved an unsafe request under switch-status or capacitor-status errors, but Twin+BDD then rejected 100\% and 57\% of the safe requests until the records are corrected, and the twin without BDD approved 44\% of unsafe requests under the switch-status error. Counting the probe stage as part of the action, probes never pushed the clean feeder over the limit and did so in 0.1\% of phase-error draws; requiring room for the smaller probe as well and escalating otherwise left this unchanged, escalated 3\% of clean requests, and raised clean false rejections from 22\% to 27\%.

\subsection{Budget and Meter Accuracy}
With one AMI read, EIV-LaR approved 19\% of unsafe requests under FDI, with four 3\% and with sixteen 0.2\%, while clean false rejections rose from 11\% to 37\%, because each extra anchored check is another chance for meter error to cross the limit; a larger budget should be paired with a small positive tolerance. Lowering the static AMI error from 0.5\% to 0.1\% reduced EIV-La's unsafe approvals under FDI from 4.0\% to 1.4\% and its false rejections from 36\% to 9\% (supplement, Fig.~S2 and Table~S7).
\space

\section{LLM Agents in the Loop}\label{sec:llm}
The gate study treats the proposer as a black box. To test whether LLM agents exhibit the same shared-evidence failure, and whether instructions can substitute for an enforced gate, we give agents seven provenance-tagged tools: retrieve the request; run state estimation; predict an action with the twin (highest predicted voltages, largest predicted rises, and the estimate and rise at any requested bus); list AMI meters (with the nearest metered bus when a bus has none); read AMI (12 reads); apply a probe (one, at most half the request); and submit a decision (approve, approve partially, reject, or escalate). The hidden evaluator scores the executed action.

\emph{Design (E2).} Four modes use the same cases for every model: (i) \emph{no gate}; (ii) \emph{protocol in prompt}, which states the EIV rules but enforces nothing; (iii) a Twin+BDD gate and (iv) the EIV gate outside the agent, evaluated on the decisions of mode (i), since the agent is never told about gates. Conditions are clean, FDI and phase-label errors, with 50 cases each (27, 29 and 23 unsafe). Under FDI two adversaries attack the same cases: one hides the violation as deeply as bad-data detection allows (target margins from 1.5\% down to 0.2\%), the other (\emph{subtle FDI}) just enough for the nominal gate to approve, so the estimate departs from the truth as little as possible; modes (i) and (ii) are repeated with an aggregator's note stating that verification is not required. References are a scripted proposer that trusts the estimate and the twin, which approves exactly what Twin+BDD approves, and a scripted policy that follows the stated protocol exactly with the same tools. We ran Claude Haiku~5.5, Sonnet~5.5 and Opus~5.5 (\texttt{claude-haiku-5-5}, \texttt{claude-sonnet-5-5}, \texttt{claude-opus-5-5}) on 4~November 2026 with default sampling and reasoning, at most 16k output tokens per call and 16 turns per episode, once per case and mode. The 1{,}500 episodes cover 150 physical cases; paired executions of the same case are compared with exact McNemar tests and case-clustered bootstrap intervals. Hypotheses: H1, without a gate agents fail like the nominal gate under FDI, because their tools expose the same estimate and twin; H2, a protocol in the prompt helps only as far as it is followed, and injected text erodes compliance; H3, an EIV gate outside the agent bounds violations regardless of the model.

\emph{H1 was not supported} (Table~\ref{tab:llm}). The trusting proposer executed every unsafe request under both FDI attackers and 13 of 23 under phase-label errors. The agents read AMI in 89--100\% of episodes without a gate; under FDI they cited readings above the estimate near the requesting DERs and trimmed, rejected or escalated the request. None approved an unsafe request in full under either attacker, and none of their decisions caused a violation (0 of 174 unsafe requests; 95\% upper bound 2.1\%); they read one of the three meters with the highest true post-action voltage in 86--93\% of unsafe FDI requests.

\emph{The check was voluntary (H2).} The injected note led Haiku to approve two unsafe FDI requests in full without reading a meter; Sonnet approved one after reading twelve meters, all at the requesting DERs rather than where the twin predicted the largest rise; Opus was not swayed. With the protocol in the prompt no model approved an unsafe FDI request in full; the one violation was a partial approval by Opus under subtle FDI that ended 0.04\% above the limit. Phase-label errors caused violations in 4 (Haiku), 1 (Sonnet) and 0 (Opus) of 23 unsafe requests without a gate, and stating the protocol changed these to 1, 4 and 2: each of those approvals passed the anchored and probe checks as written, and the scripted policy that applies the same rules exactly let 6 of 23 through, because rules simple enough to state in a prompt are coarser than the gate's statistical tests. These differences are not significant at this sample size (exact McNemar $p=0.38$ for 4 versus 1 and for 1 versus 4).

\emph{The enforced gate (H3).} Applied to the agents' decisions, the EIV gate let no unsafe request through under FDI (0 of 261) and one of 69 under phase-label errors (a partial approval that reached 1.0524~pu), and removed every violation caused by the note. Pooled over models it removed seven violations and added none (330 paired executions of 52 unsafe physical cases; $p=0.016$; difference 2.1 points, interval 0.6--4.1); per model the counts are too small for significance (Haiku 5 versus 0, $p=0.06$). Counting the probe stage, the agents' own probes exceeded 1.05~pu in 7 of the 1{,}027 episodes that probed (three physical cases, all without the protocol, at most 0.07\% above the limit); the gate's half-step probe would not have exceeded the limit for any agent decision. The agents' caution cost availability: they executed 68--76\% of the release requested in safe clean cases, mostly through partial approvals, against 96\% for the EIV gate with the trusting proposer, and adding the gate cost them at most 4 points. The 1{,}500 episodes cost US\$61 (Haiku \$1.18, Sonnet \$22.20, Opus \$37.56).

\begin{table*}[t]
\caption{LLM agents in the loop: unsafe requests that ended in a violation (\%; 29 unsafe requests per FDI cell, 23 under phase-label errors), share of the requested release executed for the 23 safe clean requests (\%), verification behavior and API cost. Gate rows apply the gate to the no-gate decisions; Covered is the share of unsafe FDI requests for which the agent read one of the three meters with the highest true post-action voltage}\label{tab:llm}
\centering\footnotesize\setlength{\tabcolsep}{4.5pt}
\begin{tabular}{@{}llrrrrrrrrr@{}}
\toprule
 & & \multicolumn{4}{c}{Violations} & Safe release & \multicolumn{3}{c}{Verification (\% of episodes)} & Cost\\
\cmidrule(lr){3-6}\cmidrule(lr){8-10}
Proposer & Mode & FDI & FDI, note & Subtle FDI & Phase & (clean) & AMI & Covered & Probe & (US\$/ep.)\\
\midrule
Scripted, trusting & No gate (= Twin+BDD) & 100 & 100 & 100 & 57 & 100 & -- & -- & -- & -- \\
 & EIV gate & 3 & 3 & 7 & 9 & 96 & -- & -- & -- & -- \\
Scripted, protocol & No gate & 0 & 0 & 0 & 26 & 87 & -- & -- & -- & -- \\
\midrule
Haiku 5.5 & No gate & 0 & 7 & 0 & 17 & 73 & 89 & 86 & 25 & 0.002 \\
 & Protocol in prompt & 0 & 0 & 0 & 4 & 71 & 100 & 93 & 61 & 0.003 \\
 & EIV gate & 0 & 0 & 0 & 4 & 73 &  &  &  &  \\
\midrule
Sonnet 5.5 & No gate & 0 & 3 & 0 & 4 & 76 & 100 & 93 & 89 & 0.047 \\
 & Protocol in prompt & 0 & 0 & 0 & 17 & 82 & 100 & 97 & 75 & 0.041 \\
 & EIV gate & 0 & 0 & 0 & 0 & 72 &  &  &  &  \\
\midrule
Opus 5.5 & No gate & 0 & 0 & 0 & 0 & 68 & 97 & 90 & 92 & 0.072 \\
 & Protocol in prompt & 0 & 0 & 3 & 9 & 59 & 100 & 90 & 69 & 0.078 \\
 & EIV gate & 0 & 0 & 0 & 0 & 68 &  &  &  &  \\
\bottomrule
\space
\end{tabular}
\end{table*}

\emph{The same checker, voluntary or required (E3).} To separate who decides to check from how the check is done, a third study gives the agents the gate itself as a tool, \texttt{verify\_action(caps)}: it runs the frozen EIV-LaR rule with the commanded-caps twin on its own budget of eight AMI reads and one probe, and issues for each cap vector it accepts a token bound to the caps and the evidence used (supplement, Section~S1). The agents keep four exploratory reads but no probe, and whole-sequence safety counts the probe's peak. In the \emph{voluntary} mode the grid executes whatever the agent submits; in the \emph{mandatory} mode, only an approval with a token for exactly those caps. Prompts differ in one sentence, and both modes run on the same fresh cases (clean, FDI with the injected note, phase-label errors; 50 cases each, two repetitions) with the same readings, so every voluntary execution has a mandatory twin. \ETresults

\begin{table}[t]
\caption{The identical checker, voluntary or mandatory (E3; 298 episodes per row): episodes that used it and executed approvals with a matching token (\%), unsafe executions ending in a violation counting the probe, safe release (\%), AMI reads and probes per episode. Scripts screen with the estimate and twin; A then submits only verified caps, B never calls the checker.}\label{tab:checker}
\centering\footnotesize\setlength{\tabcolsep}{2.3pt}
\begin{tabular}{@{}llrrrrrrrr@{}}
\toprule
Proposer & Mode & Used & Token & FDI & Phase & Clean & Release & Reads & Probes\\
\midrule
Haiku & Vol. & 93 & 89 & 1/52 & 9/50 & 0/62 & 77 & 8.0 & 0.61 \\
 & Mand. & 100 & 100 & 1/52 & 11/50 & 0/62 & 72 & 8.5 & 0.65 \\
Sonnet & Vol. & 100 & 98 & 1/52 & 10/50 & 0/62 & 81 & 10.3 & 0.72 \\
 & Mand. & 100 & 100 & 1/52 & 13/50 & 0/62 & 79 & 10.7 & 0.74 \\
Script A & Both & 68 & 100 & 1/52 & 10/50 & 0/62 & 77 & 4.6 & 0.46 \\
Script B & Vol. & 0 & 0 & 52/52 & 18/50 & 2/62 & 97 & 0.0 & 0.00 \\
 & Mand. & 0 & -- & 0/52 & 0/50 & 0/62 & 0 & 0.0 & 0.00 \\
\space
\end{tabular}
\end{table}

\section{Discussion and Limitations}
\emph{Where the gate gets its evidence matters more than how strict it is, and so does how it uses it.} A larger margin buys protection against shared-evidence failures only by rejecting most safe requests, and an adversary who knows the margin aims just inside it. Independent evidence buys the protection at much lower cost, but only if it is allowed to overrule the corrupted evidence. Fed into the estimator, four AMI readings competed with nearly critical DER rows of similar accuracy and moved the attacked estimate a sixth of the way back; used to re-check the claim directly, they stopped nearly all attacks. Down-weighting the suspect channel recovers most of that assurance, but it requires naming the suspect channel in advance, tuning a weight, recalibrating the bad-data tests and trusting a re-estimate per request, whereas the anchored check is one auditable inequality per meter. Estimation remains the better tool where the corrupted rows are few and redundant, as when only one aggregator's telemetry is false.

\emph{Independence has to cover the prediction, not only the level.} The anchored check removes the estimate's level from the safety claim but keeps the twin's predicted rise. The adaptive study showed that this input was less independent than the design rules assumed: the twin derived each unit's headroom from its reported output, so false telemetry could shrink the predicted rise, and a distortion that keeps the response linear over the probe's range escapes the probe as well. Predicting the increment from commanded set-points closed that route at no cost in clean operation; what remained were smaller distortions through nonlinear twin effects. Every input to the safety claim, including the gate's own model inputs, needs a provenance check. For agentic deployments this adds to the requirements in \cite{rafy2026twingridshield,mylonas2026governance}: the gate must obtain evidence that the agent's tools do not provide, tool outputs should carry provenance, and the gate, not the agent, should choose the corroboration points.

\emph{Agents can check, but cannot be relied on to.} Given an independent channel and the predicted rise at any node, the LLM agents compared the estimate with AMI on their own and never approved an unsafe request in full under either attacker, so exposing provenance and an independent channel in the tools is worthwhile in itself. But the check was a choice: one sentence in a data field switched it off for two of the three models, a protocol in the prompt replaced their judgment with coarser thresholds, and their own probes occasionally pushed the feeder over the limit. \ETdiscussion{} A gate outside the agent turns the check from a habit into a requirement, and costs little when the agent is already careful.

\emph{Independence is relative to the threat.} Under record errors without an adversary, DER telemetry is truthful and re-polling it is nearly as informative as AMI; under channel compromise it is worthless. AMI is not independent of everything either: its meter-to-phase map comes from the same records as the twin, so a phase-label error corrupts the twin and the interpretation of the reading together. The remaining phase-error approvals were mostly underpredictions that a small probe resolves at under two noise standard deviations; a larger probe when the readings allow it, or a second read, would buy power at the cost of excursion risk. Rejections should be treated as escalations to an operator with the failed check attached. All requests here were within 1\% of the limit; away from it the anchored check almost never rejects, so EIV can be reserved for requests with a small predicted margin. On-demand AMI reads take minutes, and the probe corresponds to a staged release, which must itself be counted against the voltage limit.

\emph{Limitations.} The study uses one feeder and steady-state snapshots, and all near-limit requests arose downstream of regulator~4, which hosts half the PV capacity. The adversaries are white-box but bounded: the adaptive attacker searched a declared pool of linearized attack families with a fixed budget, and its validation results hold for that search, not for every possible attack. The stability of the predicted rise is a local property of the twin around the estimated state, and the probe's evidence covers only the direction of change actually applied. Operating points calibrated on development cases transferred imperfectly (15--18\% instead of 10\% false rejections on held-out cases, equally for every gate). AMI errors are Gaussian static and per-read terms without latency, and the estimator is our own implementation rather than PMDSE \cite{vanin2020pmdse}. The LLM studies use one model family, 50 cases per cell and tools that we designed, and their attackers do not adapt to the agents' checks. Time series, larger feeders such as the IEEE 8500-node system, communication-network emulation and hardware-in-the-loop inverters are natural next steps.
\space

\section{Conclusion}
Physics-based gates are becoming the safety layer of agentic AI in power systems, but a gate that evaluates an action against the same evidence the agent used inherits that evidence's failures. On a hidden-truth IEEE 123-bus testbed, stealthy false data on DER telemetry made an AC twin gate with bad-data detection approve \UARfdiBDD{}\% of unsafe export requests, and safety margins large enough to help rejected nearly all safe requests. Re-establishing the safety claim at the action's footprint with independent AMI evidence, choosing corroboration points from the predicted rise rather than the estimate, and testing the model's response with a bounded probe reduced unsafe approvals to \UARfdiEIV{}\% under attack. The same readings absorbed into the state estimator bought far less assurance, and attacks adapted to the gate showed that independence must extend to the twin's predicted response, which is best computed from the operator's own set-points. LLM agents that could see independent evidence often checked on their own, but an injected note or a model error could stop them; offered the gate as a tool, they used it almost always, and only enforcing it made the check unconditional. Agreement between an agent and its checker is assurance only when the checker can see something the agent cannot.
\space

\bibliographystyle{IEEEtran}
\bibliography{refs}

\end{document}